# Effect of decorating NiO nanoparticles on superconducting properties of YBCO

D.M. Gokhfeld[1], S.V. Semenov[1], M.I. Petrov[1], I.V. Nemtsev[1,2], M.S. Molokeev[1], V.L. Kirillov[3], O.N. Martyanov[3]

[1] Kirensky Institute of Physics, Federal Research Center KSC SB RAS, Krasnoyarsk, 660036 Russian Federation

[2] Federal Research Center KSC SB RAS, Krasnoyarsk, 660036 Russian Federation

[3] Boreskov Institute of Catalysis SB RAS, Novosibirsk, 630090 Russian Federation

E-mail: gokhfeld@iph.krasn.ru

The influence of adding 23 nm NiO nanoparticles on the magnetic hysteresis loops and critical current density of the high-temperature superconductor $YBa_2Cu_3O_{7-\delta}$ has been investigated. The samples were prepared using a fast annealing method that prevents chemical interaction between the components and does not reduce the critical temperature of the superconductor. Compared to the undoped YBCO sample, the critical current density increases in the samples doped with NiO nanoparticles in magnetic fields greater than 6 kOe. The sample containing 0.5 weight percent of NiO nanoparticles exhibits the greatest enhancement in critical current density at 10 kOe (1.36 times greater than the undoped sample).

## 1. Introduction

The influence of various defects on an increase in the critical current density of superconductors is an important topic in the solid-state physics and materials science [1–3]. According to theoretical works [4–7], magnetic inclusions serve as particularly effective pinning centers. A number of experimental studies have found a significant increase in the critical current density $J_c$ upon the incorporation of magnetic nanoparticles into superconductors [8–13]. However, when adding magnetic nanoparticles to high-$T_c$ superconductors, it is difficult to avoid chemical interaction leading to material degradation [14,15]. One method to reduce chemical interaction is surface decoration [16–18], where magnetic nanoparticles are placed on the surface of the superconductor.

Nickel (II) oxide NiO nanoparticles can be promising candidate for pinning improvement in REBCO superconductors. Study [19] found that adding up to 0.05 wt.% of 24-nm-average-size NiO nanoparticles in $YBa_2Cu_3O_{7-\delta}$ (YBCO) increased $J_c$. The $J_c$ values were determined from transport measurements. In a recent work [20], the addition of up to 0.2 wt.% 100-nm-average-size NiO nanoparticles to YBCO increased $J_c$ in low fields. The $J_c$ values were determined from magnetic measurements. In these works [19,20], YBCO samples were prepared using solid-state synthesis, and the addition of NiO nanoparticles occurred at the final stage with subsequent 12-hour annealing at a temperature of about 900 °C. With this production method, Ni incorporation into the YBCO crystal lattice occurs [21,22].

The aim of the present work is to investigate the effect of 23 nm NiO nanoparticles on the magnetization and critical current density of YBCO. The fast annealing technique [23] is used to obtain the NiO doped YBCO samples. Instead works [21,22], we try to decorate the surface of superconducting granules with magnetic nanoparticles and to avoid the incorporation of Ni into the YBCO crystal lattice.

## 2. Method

A series of samples with different nanoparticle content from 0.1 to 6 wt.% was prepared. The initial $YBa_2Cu_3O_{7-\delta}$ material was prepared by standard solid-state synthesis technology from $Y_2O_3$, BaO, and CuO powders. NiO nanoparticles with an average size of 23 nm were prepared at Institute of Catalysis SB RAS [24,25].

A required amount of NiO nanoparticles was added to the initial YBCO material. A tenfold amount of alcohol was added to the powder mixture, and the alcohol suspension was stirred. Then the alcohol was evaporated with mild heating on a hotplate. The dry residue was thoroughly ground in a mortar and subjected to hydrostatic pressing (100 MPa). Next, fast annealing was performed [23]: the pressed materials were placed for 2 minutes in a furnace heated to a temperature of 920 °C. Then they were moved to a furnace heated to a temperature of 400 °C and held at this temperature for 10 hours.

X-ray phase analysis of the synthesized materials was performed on a Haoyuan DX-2700BH powder diffractometer. Micrographs of the surface of samples were obtained using a Hitachi TM4000Plus scanning electron microscope. The distribution of chemical elements on the sample surface was determined using a Bruker XFlash 630Hc energy dispersive spectrometer. Magnetization measurements were carried out on a Lakeshore VSM 8604 vibrating sample magnetometer.

## 3. Results and Discussion

The X-ray phase analysis showed only the appearance of the NiO phase in the YBCO samples with added nanoparticles. The average size of the YBCO granules, determined from scanning electron microscopy images, is about 2 μm (Fig. 1a). The energy dispersive spectroscopy showed a uniform distribution of NiO on the surface of the granules (Fig. 1b).

Figure 2 shows the temperature dependences of magnetization, measured during heating in a field of 100 Oe after zero-field cooling. For all samples, the magnetization reaches a value of M ≈ 0 at a temperature of 93.0 K.

The measured magnetic hysteresis loops of all samples are shown in Fig. 3. The upper inset in Fig. 3 shows an enlarged fragment of the magnetic hysteresis loops near the magnetization maximum. With increasing NiO content $x > 0.005$, the maximum signal values monotonically decrease.

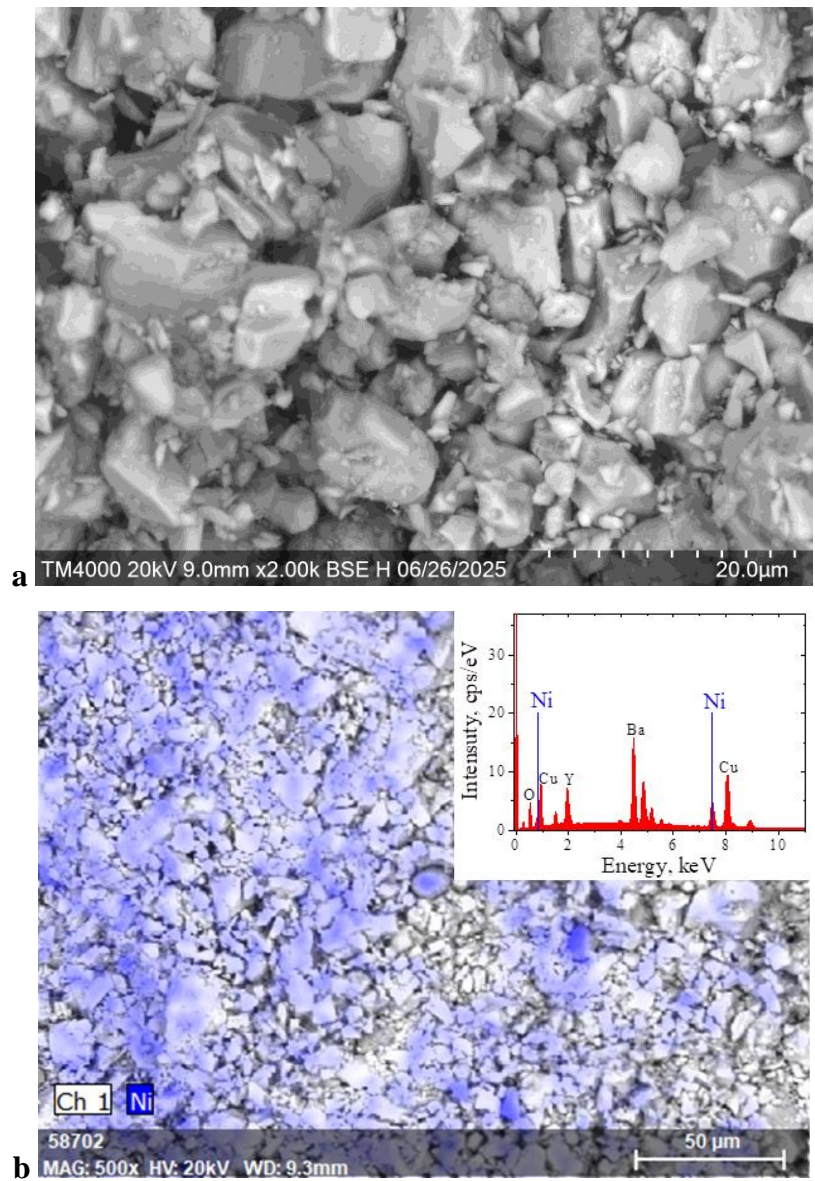

Fig. 1. Micrograph of the sample with $x = 0.06$ (a) and distribution of Ni on the surface of this sample (b). The insert shows the energy dispersive spectrum. Blue marks are for the peaks corresponding to Ni.

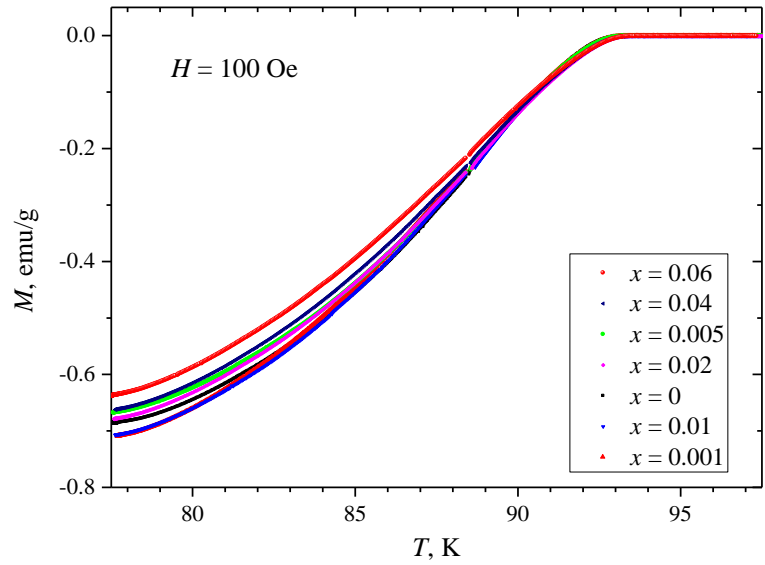

Fig. 2. Temperature dependencies of magnetization at $H$ = 100 Oe (zero field cooling).

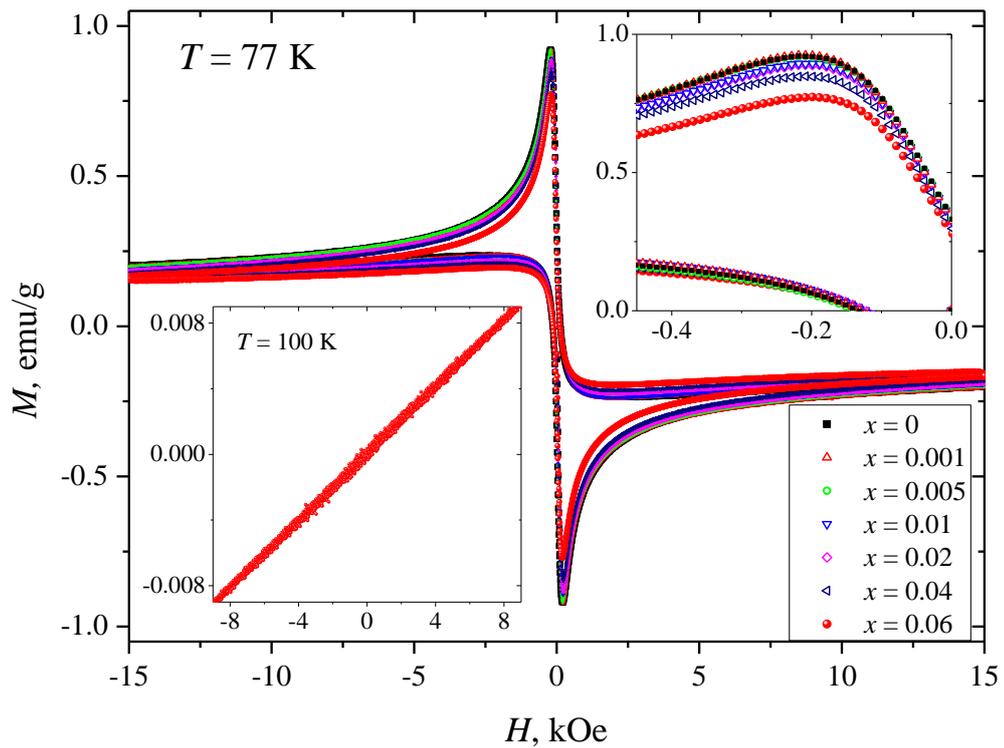

Fig. 3. Magnetic hysteresis loops at $T$ = 77 K. The upper inset shows an enlarged fragment of these loops near the magnetization maximum. The lower inset shows the $M(H)$ dependence for $x$ = 0.06 at $T$ = 100 K.

The magnetic subsystem of NiO can also contribute to the magnetic hysteresis loop. To account for this contribution, we measured the $M(H)$ dependence for the sample with $x = 0.06$ at $T = 100$ K (lower inset Fig. 3), i.e., above the critical temperature of YBCO. This dependence shows no hysteresis. It can be assumed that at 77 K there is also no contribution from the magnetic subsystem to the distance $\Delta M$ between the upper and lower branch of a magnetic hysteresis loop.

The critical current density was calculated from the magnetic hysteresis loops using the formula $J_c(H) = 3\Delta M(H)/D$, where $D$ is the characteristic scale. For the investigated samples, the size $D$ corresponds to the granule size [26,27], because the magnetization of polycrystalline YBCO is determined by the magnetic flux trapped by the granules, and the contribution of intergranular currents to the magnetization is negligible [28]. The $J_c$ values were estimated for $D = 2.2$ μm, as in work [29]. Figure 4 shows the calculated $J_c(H)$ dependences at $T = 77$ K. The obtained values of the intragranular critical current density are close to the $J_c$ of REBCO tapes [11,16].

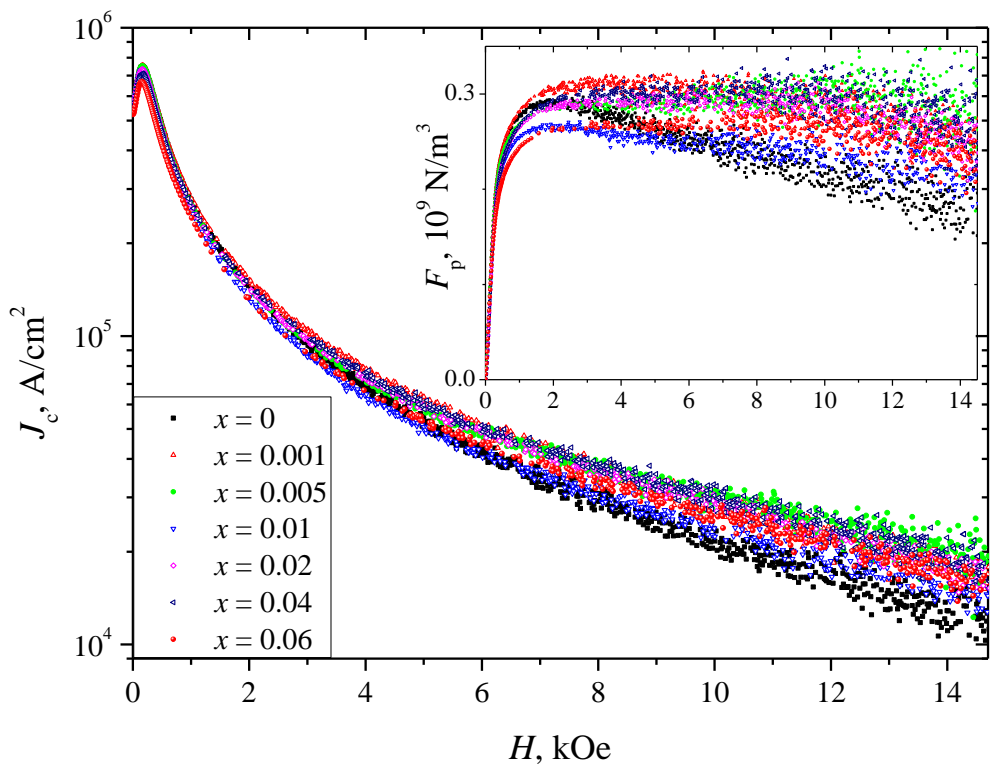

Fig. 4. The intragranular critical current density at $T = 77$ K. Inset shows the pinning force density versus magnetic field.

Near $H = 0$, low concentrations of NiO ($x = 0.001, 0.005$) have practically no effect on $J_c$, while an increase in nanoparticle content leads to a decrease in $J_c$. However, in magnetic fields greater than 6 kOe, the $J_c$ values in all samples with NiO nanoparticles become greater than in the undoped sample with $x = 0$. The dependencies of $J_c$ on $x$ in fields of 160 Oe and 10 kOe are shown in Fig. 5. The $J_c$ values in Fig. 5 are normalized to the $J_c$ values of the undoped sample, $J_c(x = 0) = 0.75$ MA/cm$^2$ at 160 Oe, $J_c(x = 0) = 0.22$ MA/cm$^2$ at 10 kOe. At 10 kOe, a maximum increase in $J_c$ by a factor of 1.36 was found in the sample with $x = 0.005$.

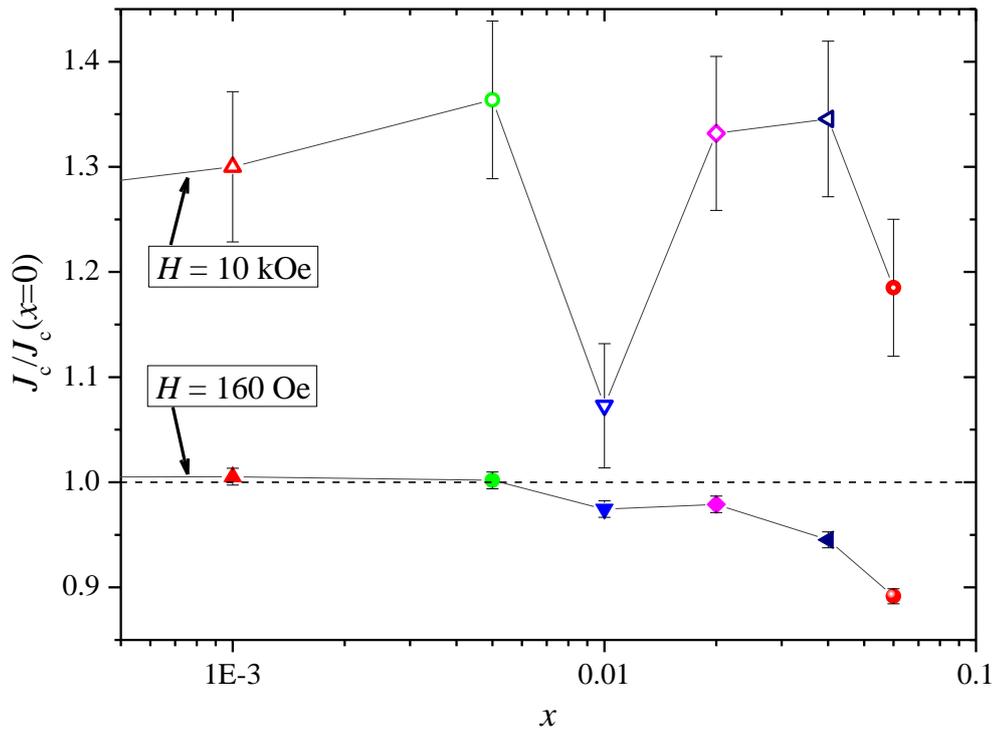

Fig. 5. Critical current density at $T = 77$ K normalized to the $J_c$ values of the undoped YBCO. Full symbols are the normalized current density at 160 Oe, empty symbols are the normalized current density at 10 kOe. Points above the dashed line correspond to $J_c$ values greater than $J_c$ of the undoped sample in the corresponding field.

The pinning force $F_p = \mu_0 H J_c$ in the undoped YBCO decreases faster than in the samples with NiO nanoparticles (inset in Fig. 4). Consequently, in the NiO doped samples, the value of the irreversibility field, at which $F_p = 0$, should be greater than in the undoped YBCO.

In work [30], the addition of NiO nanoparticles to YBCO led to the formation of pinning centers with a size of about 10 nm. In these samples, an increase in $J_c$ in high fields was also observed compared to the undoped YBCO.

The obtained data show that NiO nanoparticles improve the intragrain critical current density. This $J_c$ enhancement is due to an increase of surface pinning by magnetic nanoparticles accommodated on the surface of superconducting granules.

## 4. Conclusion

The NiO doped YBCO samples have been prepared using the fast annealing technique. NiO nanoparticles (23 nm) have been uniformly distributed on the surface of the YBCO grains without chemical degradation of the superconducting matrix. This is a significant improvement over traditional long-duration sintering processes, which often lead to the incorporation of Ni into the YBCO crystal lattice, degrading its superconducting properties.

It has been established that adding magnetic NiO nanoparticles increases the critical current density $J_c$ in high magnetic fields. While high concentrations of NiO nanoparticles (>0.5 wt.%) are detrimental to $J_c$ in low magnetic fields, all doped samples up to 6 wt.% exhibit a substantial increase in $J_c$ in applied magnetic fields exceeding 6 kOe. The optimal doping level was found to be 0.5 wt.% NiO. This sample has 1.36-fold increase in $J_c$ at 10 kOe compared to the undoped YBCO.

**Acknowledgements** The measurements were carried out on the equipment of Krasnoyarsk Regional Center of Research Equipment of Federal Research Center «Krasnoyarsk Science Center SB RAS». We would like to thank T.D. Balaev and S.E. Nikitin for their assistance with the data analysis.

**Funding** This study was supported by grant 24-22-00053 provided by the Russian Science Foundation, https://rscf.ru/project/24-22-00053/